\documentclass[notitlepage,nofootinbib,preprintnumbers,amssymb,superscriptaddress]{revtex4-1}
\usepackage{amsfonts,amssymb,amsmath,graphicx,color,bm}
\usepackage[linktocpage=true]{hyperref}
\hypersetup{
colorlinks=true,
citecolor=red,
linkcolor=blue,
urlcolor=blue,
}

\begin{document}

\title{Bound gravitational waves in a dielectric medium and a constant magnetic field}

\author{A.N. Morozov}
\email[E-mail:]{amor59@mail.ru, amor@bmstu.ru}
\affiliation{Department of Physics, Bauman Moscow State Technical University,
  Moscow, 105005, Russia}

\author{V.I. Pustovoit}
\email[E-mail:]{vladpustovoit@gmail.com}
\affiliation{Department of Physics, Bauman Moscow State Technical University,
  Moscow, 105005, Russia}
\affiliation{Scientific and Technological Centre of Unique Instrumentation of the Russian Academy of Sciences, Moscow, 117342, Russia}

\author{I.V. Fomin}
\email[E-mail:]{ingvor@inbox.ru}
\affiliation{Department of Physics, Bauman Moscow State Technical University,
  Moscow, 105005, Russia}


\begin{abstract}
A description is made of the process of excitation of bound longitudinal-transverse gravitational waves during the propagation of electromagnetic waves in a dielectric medium. It is shown that the speed of such gravitational waves is less than the speed of light in a vacuum and coincides with the speed of an electromagnetic wave in matter. A description of the propagation of a bound gravitational waves in a dielectric in the presence of a constant magnetic field is suggested as well. It is claimed that these gravitational waves in a dielectric medium are forced ones and they cannot exist in a free state.
\end{abstract}

\maketitle

\section{Introduction}

The problem of describing the gravitational field induced by electromagnetic waves has been considered previously in a fairly large number of works \cite{Tolman:1931zza,Hegarty1969,Bonnor1969,Aichelburg:1970dh,Nackoney1973,Voronov:1974,Scully:1979xe,Dray:1984ha,Bonnor:2009zza,
vanHolten:2008ts,Ratzel:2015nqf,vanHolten:2018izl}. Also, the excitation of gravitational waves when a strong electromagnetic wave propagates in empty space is considered in the works \cite{Eddington1,Gertsenshtein1962,Grishchuk:1973qz,Grishchuk:1975,Pustovoit:1981za,Nikishov:2010zz,Morozov2020,Morozov:2020snv}. In these works, it was shown that an electromagnetic wave excites a gravitational wave having a propagation velocity that coincides with the speed of an electromagnetic wave in a vacuum. The equality of the speeds of electromagnetic and gravitational waves in empty space is confirmed experimentally \cite{Abbott:2016blz,Abbott:2016nmj,Aasi:2013wya,Monitor:2017mdv}.

Also, in \cite{Zel'dovich1974,Gerlach:1974zz,Zeldovich1983,Raffelt:1987im,Fargion:1995mm,
Marklund:1999sp,Dolgov:2012be,Kolosnitsyn:2015zua,Dolgov:2017bpj,Ejlli:2018hke} it was shown that when an electromagnetic wave propagates in a strong constant magnetic field, a transverse gravitational wave with a linearly increasing amplitude increases from distance. Moreover, the speed of such a gravitational wave also coincides with the speed of light in a vacuum.
Thus, the analysis of generation of gravitational waves by standing electromagnetic waves in resonators with a dielectric medium \cite{Grishchuk:1975,Nikishov:2010zz,Denisov1977} turns out to be limited to the case of excitation of transverse gravitational waves propagating with the speed of light in empty space.

The aim of this work is to describe the bound longitudinal-transverse gravitational waves which induced by
a strong electromagnetic wave in a dielectric medium in the presence of a strong constant magnetic field and
without such a field. A analysis of the longitudinal-transverse gravitational waves coupled with an electromagnetic wave in vacuum was presented earlier in \cite{Morozov2020,Morozov:2020snv}.

To consider a coupled plane gravitational wave in the direction, we will use the Einstein-Maxwell equations which describe the gravitational field created by the electromagnetic field \cite{Landau,Weber,Maggiore:2018sht}
\begin{equation}
\label{EEQ}
R_{\mu\nu}=\frac{8\pi G}{c^{4}}\left(T_{\mu\nu}-\frac{1}{2}g_{\mu\nu}T\right),
\end{equation}
where $G$ is the gravitational constant, $c$ is the velocity of light in vacuum, $T=g^{\mu\nu}T_{\mu\nu}$ and $\mu,\nu=0,1,2,3$.

One can also write equation (\ref{EEQ}) as
\begin{equation}
\label{EEQ1}
R_{\mu\nu}=\frac{8\pi G}{c^{4}}\tilde{T}_{\mu\nu},
\end{equation}
where the reduced energy-momentum tensor $\tilde{T}_{\mu\nu}$ is
\begin{equation}
\label{TC}
\tilde{T}_{\mu\nu}=T_{\mu\nu}-\frac{1}{2}g_{\mu\nu}T.
\end{equation}

For weak gravitational fields, the space-time metric $g_{\mu\nu}$ is reduced to the case of Minkowski space-time with a small perturbations $g_{\mu\nu}=\eta_{\mu\nu}+h_{\mu\nu}$, where $|h_{\mu\nu}|\ll 1$ \cite{Landau,Weber,Maggiore:2018sht}.
Thus, the basis of the analysis is the direct solution of equations (\ref{EEQ1}) for the perturbed Minkowski space-time and reduced energy-momentum tensor of the electromagnetic wave (\ref{TC}), taking into account the contribution of the dielectric medium.

The article is organized as follows:  in section \ref{sect2} the energy-momentum tensor of the electromagnetic field is determined taking into account the dielectric medium and the case of a linear dependence of the permittivity on the density of matter is considered.
Section \ref{sect3} presents the Einstein-Maxwell equations for an electromagnetic wave passing through a dielectric.
In Sec. \ref{sect4} the gravitational-wave solutions of the Einstein-Maxwell equations corresponding to an electromagnetic wave in dielectric media are presented. It is noted that these solutions were obtained beyond the harmonic gauge. However, with the passage to the limit to empty space, the harmonic gauge is satisfied.
The components of the energy-momentum tensor of electromagnetic wave in the presence of a constant magnetic field are defined in Sec. \ref{sect5}.
Based on this energy-momentum tensor, in  Sec. \ref{sect6} the solutions that describe bound gravitational waves in a dielectric medium are obtained. It is also noted that these solutions for transverse gravitational-wave components in a dielectric media generalize the known ones in empty space.
In conclusion, the results of this study are discussed.

\section{The energy-momentum tensor for propagating electromagnetic wave in a dielectric medium}\label{sect2}

Let a plane harmonic electromagnetic wave be emitted from a point $x=0$ in the direction of the axis $x^{1}=x$ in a medium with dielectric constant $\varepsilon$ and magnetic one $\mu=1$.

The expressions for the electric and magnetic fields for such a wave are
\begin{eqnarray}
\label{EF}
&&E_{y}=E_{0}\cos\left(\omega\left(t-\frac{nx}{c}\right)\right),\\
\label{MF}
&&H_{z}=H_{0}\cos\left(\omega\left(t-\frac{nx}{c}\right)\right),
\end{eqnarray}
where the refractive index of the medium $n=\sqrt{\varepsilon}$, and the amplitudes of the electric $E_{0}$ and magnetic $H_{0}$ fields are related as
\begin{equation}
H_{0}=\sqrt{\varepsilon}E_{0}=nE_{0}.
\end{equation}

The nonzero components of the energy-momentum tensor $T_{\mu\nu}$ for this case are \cite{LandauECM}
\begin{eqnarray}
\label{Energy}
&&T_{00}=\frac{1}{8\pi}\left(\varepsilon E^{2}_{y}+H^{2}_{z}\right)=
\frac{n^{2}E^{2}_{0}}{8\pi}\left(1+\cos\left(2\omega\left(t-\frac{nx}{c}\right)\right)\right),\\
\label{Pointing}
&&T_{01}=T_{10}=-\frac{1}{4\pi}\varepsilon E_{y}H_{z}=
-\frac{nE^{2}_{0}}{8\pi}\left(1+\cos\left(2\omega\left(t-\frac{nx}{c}\right)\right)\right),\\
\label{Maxvell}
&&T_{\alpha\alpha}=\frac{1}{4\pi}\left[-\varepsilon E_{\alpha}E_{\alpha}-
H_{\alpha}H_{\alpha}+\frac{1}{2}\left(\varepsilon-\rho\left(\frac{\partial\varepsilon}{\partial\rho}\right)_{T_{0}}\right)
\left(E^{2}_{y}+\frac{H^{2}_{z}}{\varepsilon}\right)\right],
\end{eqnarray}
where $\alpha=1,2,3$.

The last expression (\ref{Maxvell}) demonstrates the fact that not only the electromagnetic field, but also the dielectric medium contributes to the stress tensor $T_{\alpha\alpha}$.
Also, in the expression (\ref{Maxvell}), it is taken into account that when calculating the tensor $T_{\alpha\alpha}$, it is necessary to consider the dependence of the dielectric constant $\varepsilon$ from the density of matter $\rho$ for a constant temperature $T_{0}$ \cite{LandauECM}.

For gases, this dependence has a simple form
\begin{equation}
\label{vareps}
\varepsilon=1+\kappa\rho,
\end{equation}
where $\kappa$ is a constant depending on the chemical composition of a gas.

If the dependence of the dielectric constant of a medium on its density satisfies relation (\ref{vareps}), then one can obtain following expressions for the components of the energy-momentum tensor
\begin{eqnarray}
\label{T11}
&&T_{11}=\frac{E^{2}_{0}}{8\pi}\left(1+\cos\left(2\omega\left(t-\frac{nx}{c}\right)\right)\right),\\
\label{T22}
&&T_{22}=T_{33}=\frac{(1-n^{2})E^{2}_{0}}{8\pi}\left(1+\cos\left(2\omega\left(t-\frac{nx}{c}\right)\right)\right),
\end{eqnarray}

Also, we note that expressions (\ref{Energy}), (\ref{Pointing}), (\ref{T11}) and (\ref{T22}) for the components of the energy-momentum tensor satisfy the conservation laws
\begin{equation}
\label{CL}
\frac{\partial T^{\mu\nu}}{\partial x^{\nu}}=
\frac{\partial(\eta^{\mu\sigma}\eta^{\nu\lambda}T_{\sigma\lambda})}{\partial x^{\nu}}=0.
\end{equation}

Formulas (\ref{Energy}), (\ref{T11}) and (\ref{T22}) allow one to find the expression for the scalar $T$ in the case of a small curvature of space-time
\begin{equation}
\label{ScalarT}
T=\eta_{\mu\nu}T^{\mu\nu}=T_{00}-T_{11}-T_{22}-T_{33}=
\frac{3(n^{2}-1)E^{2}_{0}}{8\pi}\left(1+\cos\left(2\omega\left(t-\frac{nx}{c}\right)\right)\right).
\end{equation}

Then, using expressions (\ref{TC}), (\ref{Energy}), (\ref{T11}), (\ref{T22}) and (\ref{ScalarT}), we obtain
\begin{eqnarray}
\label{T00S}
&&\tilde{T}_{00}=\frac{(3-n^{2})E^{2}_{0}}{16\pi}\left(1+\cos\left(2\omega\left(t-\frac{nx}{c}\right)\right)\right),\\
\label{T11S}
&&\tilde{T}_{11}=\frac{(3n^{2}-1)E^{2}_{0}}{16\pi}\left(1+\cos\left(2\omega\left(t-\frac{nx}{c}\right)\right)\right),\\
\label{T22S}
&&\tilde{T}_{22}=\tilde{T}_{33}=\frac{(n^{2}-1)E^{2}_{0}}{16\pi}
\left(1+\cos\left(2\omega\left(t-\frac{nx}{c}\right)\right)\right),
\end{eqnarray}
also, taking into account the expression (\ref{Pointing}), one has $\tilde{T}_{01}=\tilde{T}_{10}=T_{01}=T_{10}$.

\section{The Einstein-Maxwell equations for an electromagnetic wave in a dielectric medium}\label{sect3}

When describing the propagation of a plane gravitational wave in direction $x^{1}=x$,
the Einstein-Maxwell equations have the following form \cite{Morozov2020}
\begin{eqnarray}
\label{E1}
&&-\frac{\partial^{2} h_{11}}{c^{2}\partial t^{2}}+2\frac{\partial^{2} h_{01}}{c\partial t\partial x}-
\frac{\partial^{2} h_{00}}{\partial x^{2}}+\frac{\partial^{2} h_{22}}{c^{2}\partial t^{2}}+
\frac{\partial^{2} h_{33}}{c^{2}\partial t^{2}}=\frac{16\pi G}{c^{4}}\tilde{T}_{00},\\
\label{E2}
&&\frac{\partial^{2} h_{11}}{c^{2}\partial t^{2}}-2\frac{\partial^{2} h_{01}}{c\partial t\partial x}+
\frac{\partial^{2} h_{00}}{\partial x^{2}}+\frac{\partial^{2} h_{22}}{\partial x^{2}}+
\frac{\partial^{2} h_{33}}{\partial x^{2}}=\frac{16\pi G}{c^{4}}\tilde{T}_{11},\\
\label{E3}
&&\frac{\partial^{2} h_{22}}{c\partial t\partial x}+\frac{\partial^{2} h_{33}}{c\partial t\partial x}
=\frac{16\pi G}{c^{4}}\tilde{T}_{01},\\
\label{E4}
&&\frac{\partial^{2} h_{22}}{\partial x^{2}}-\frac{\partial^{2} h_{22}}{c^{2}\partial t^{2}}
=\frac{16\pi G}{c^{4}}\tilde{T}_{22},\\
\label{E5}
&&\frac{\partial^{2} h_{33}}{\partial x^{2}}-\frac{\partial^{2} h_{33}}{c^{2}\partial t^{2}}
=\frac{16\pi G}{c^{4}}\tilde{T}_{33},
\end{eqnarray}
where the condition $h_{01}=h_{10}$ was used.

Addition of (\ref{E1}) and (\ref{E2}) gives
\begin{equation}
\label{E6}
\frac{\partial^{2} \tilde{h}}{c^{2}\partial t^{2}}+\frac{\partial^{2} \tilde{h}}{\partial x^{2}}=
\frac{16\pi G}{c^{4}}(\tilde{T}_{00}+\tilde{T}_{11}),
\end{equation}
where
\begin{equation}
\label{E8}
\tilde{h}= h_{22}+h_{33}.
\end{equation}

The system of equations (\ref{E3} )--(\ref{E6}), taking into account the notation (\ref{E8}), can be noted in  following form
\begin{eqnarray}
\label{E9}
&&\frac{\partial^{2} \tilde{h}}{c\partial t\partial x}= - \frac{2nGE^{2}_{0}}{{{c^4}}}\left( {1 + \cos \left( {2\omega \left( {t - \frac{{nx}}{c}} \right)} \right)} \right),\\
\label{E10}
&&\frac{{{\partial ^2}\tilde h}}{{\partial {x^2}}} + \frac{{{\partial ^2}\tilde h}}{{{c^2}\partial {t^2}}} = \frac{{2\left( {{n^2} + 1} \right)GE_0^2}}{{{c^4}}}\left( {1 + \cos \left( {2\omega \left( {t - \frac{{nx}}{c}} \right)} \right)} \right),\\
\label{E11}
&&\frac{{{\partial ^2}{h_{22}}}}{{\partial {x^2}}} - \frac{{{\partial ^2}{h_{22}}}}{{{c^2}\partial {t^2}}} = \frac{{\left( {{n^2} - 1} \right)GE_0^2}}{{{c^4}}}\left( {1 + \cos \left( {2\omega \left( {t - \frac{{nx}}{c}} \right)} \right)} \right),\\
\label{E12}
&&\frac{{{\partial ^2}{h_{33}}}}{{\partial {x^2}}} - \frac{{{\partial ^2}{h_{33}}}}{{{c^2}\partial {t^2}}} = \frac{{\left( {{n^2} - 1} \right)GE_0^2}}{{{c^4}}}\left( {1 + \cos \left( {2\omega \left( {t - \frac{{nx}}{c}} \right)} \right)} \right),
\end{eqnarray}
which allows one to obtain the components $h_{22}$, $h_{33}$ and function $\tilde{h}$.

\section{Bound gravitational wave in a dielectric medium}\label{sect4}

When taking into account only the components that describe the gravitational wave with frequency $2\omega$, from
equations (\ref{E9})--(\ref{E12}), one has the following solutions
\begin{eqnarray}
\label{S1}
&&{h_{22}} = {h_{33}} =  - \frac{{GE_0^2}}{{4{c^2}{\omega ^2}}}\cos \left( {2\omega \left( {t - \frac{{nx}}{c}} \right)} \right),\\
\label{S2}
&&\tilde h = {h_{22}} + {h_{33}} =  - \frac{{GE_0^2}}{{2{c^2}{\omega ^2}}}\cos \left( {2\omega \left( {t - \frac{{nx}}{c}} \right)} \right).
\end{eqnarray}

Substituting solution (\ref{S2}) into equation (\ref{E2}) taking into account (\ref{T11S}) allows one to write the equation for the components $h_{00}$, $h_{01}=h_{10}$ and $h_{11}$. Taking into account only the wave components of the coupled gravitational wave for the frequency $2\omega$, one has
\begin{equation}
\label{EM1}
\frac{{{\partial ^2}{h_{11}}}}{{{c^2}\partial {t^2}}} - 2\frac{{{\partial ^2}{h_{01}}}}{{c\partial t\partial x}} + \frac{{{\partial ^2}{h_{00}}}}{{\partial {x^2}}} = \frac{{\left( {{n^2} - 1} \right)GE_0^2}}{{{c^4}}}\cos \left( {2\omega \left( {t - \frac{{nx}}{c}} \right)} \right).
\end{equation}

The solution to this equation has the following form
\begin{eqnarray}
\label{S3}
&&{h_{00}} =  - \frac{{GE_0^2}}{{2{c^3}\omega }}x\sin \left( {2\omega \left( {t - \frac{{nx}}{c}} \right)} \right) + \frac{{GE_0^2}}{{4{c^2}{\omega ^2}}}\cos \left( {2\omega \left( {t - \frac{{nx}}{c}} \right)} \right),\\
\label{S4}
&&{h_{01}} = {h_{10}} = \frac{{nGE_0^2}}{{2{c^3}\omega }}x\sin \left( {2\omega \left( {t - \frac{{nx}}{c}} \right)} \right) - \frac{{nGE_0^2}}{{4{c^2}{\omega ^2}}}\cos \left( {2\omega \left( {t - \frac{{nx}}{c}} \right)} \right),\\
\label{S5}
&&{h_{11}} =  - \frac{{{n^2}GE_0^2}}{{2{c^3}\omega }}x\sin \left( {2\omega \left( {t - \frac{{nx}}{c}} \right)} \right) + \frac{{GE_0^2}}{{4{c^2}{\omega ^2}}}\cos \left( {2\omega \left( {t - \frac{{nx}}{c}} \right)} \right).
\end{eqnarray}

From the obtained solutions (\ref{S1}) and (\ref{S3})--(\ref{S5}) it follows that the speed of a coupled gravitational wave propagating in a medium with a dielectric is less than the speed of light in empty space and is equal to the speed of an electromagnetic wave $v_{g}=v_{e}=c/n$.

The gravity wave energy flux density $ct^{01}$ can be calculated by the formula
\begin{equation}
\label{EnFlux}
ct^{01}=-\frac{c^{5}}{32\pi G}\left(\frac{\partial h_{00}}{\partial x^{0}}\frac{\partial h_{00}}{\partial x^{1}}-
\frac{\partial h_{10}}{\partial x^{0}}\frac{\partial h_{01}}{\partial x^{1}}-
\frac{\partial h_{01}}{\partial x^{0}}\frac{\partial h_{10}}{\partial x^{1}}+\frac{\partial h_{11}}{\partial x^{0}}\frac{\partial h_{11}}{\partial x^{1}}+\frac{\partial h_{22}}{\partial x^{0}}\frac{\partial h_{22}}{\partial x^{1}}+\frac{\partial h_{33}}{\partial x^{0}}\frac{\partial h_{33}}{\partial x^{1}}\right).
\end{equation}

Substitution of expressions (\ref{S1}) and (\ref{S3})--(\ref{S5}) into formula (\ref{EnFlux}), taking into account the condition
$x\gg\frac{\omega}{c}$, gives that, as a first approximation, one can obtain an expression for the energy flux density of a gravitational wave as
\begin{equation}
\label{EnFlux1}
c{t^{01}} = \frac{{n{{\left( {{n^2} - 1} \right)}^2}GE_0^4}}{{32\pi {c^3}}}{x^2}{\cos ^2}\left( {2\omega \left( {t - \frac{{nx}}{c}} \right)} \right).
\end{equation}

We note that for $n\rightarrow1$ the energy flux density of the gravitational wave becomes equal to zero $ct^{01}=0$.

It should be noted that solutions (\ref{S1}) and (\ref{S3})--(\ref{S5}) don't satisfy the requirement of harmonic gauge  \cite{Landau,Weber,Maggiore:2018sht}
\begin{equation}
\label{HC}
\frac{\partial}{\partial x^{\nu}}\left(\eta^{\nu\sigma}h_{\sigma\mu}-\frac{1}{2}\delta^{\nu}_{\mu}h\right)=0,
\end{equation}
where
\begin{equation}
\label{HCA}
h=h_{00}-h_{11}-h_{22}-h_{33}=
\frac{{\left( {{n^2} - 1} \right)GE_0^2}}{{2{c^3}\omega }}x\sin \left( {2\omega \left( {t - \frac{{nx}}{c}} \right)} \right) - \frac{{GE_0^2}}{{2{c^2}{\omega ^2}}}\cos \left( {2\omega \left( {t - \frac{{nx}}{c}} \right)} \right).
\end{equation}

Then from expression (\ref{HC}), taking into account (\ref{HCA}), one can obtain
\begin{eqnarray}
\label{HC2}
&&\frac{\partial }{{c\partial t}}\left( {{h_{00}} - \frac{1}{2}h} \right) - \frac{{\partial {h_{10}}}}{{\partial x}} = \frac{{\left( {{n^2} - 1} \right)GE_0^2}}{{2{c^4}}}x\cos \left( {2\omega \left( {t - \frac{{nx}}{c}} \right)} \right) + \frac{{n\left( {n - 1} \right)GE_0^2}}{{2{c^3}\omega }}\sin \left( {2\omega \left( {t - \frac{{nx}}{c}} \right)} \right),\\
\label{HC3}
&&\frac{{\partial {h_{01}}}}{{c\partial t}} - \frac{\partial }{{\partial x}}\left( {{h_{11}} + \frac{1}{2}h} \right) = \frac{{n\left( {1 - {n^2}} \right)GE_0^2}}{{2{c^4}}}x\cos \left( {2\omega \left( {t - \frac{{nx}}{c}} \right)} \right) + \frac{{{{\left( {n - 1} \right)}^2}GE_0^2}}{{2{c^3}\omega }}\sin \left( {2\omega \left( {t - \frac{{nx}}{c}} \right)} \right).
\end{eqnarray}

Since the right-hand sides of these expressions are not equal to zero, harmonic gauge (\ref{HC}) for solutions (\ref{S1}) and (\ref{S3})--(\ref{S5}) is violated. For the case $n\rightarrow1$, the right-hand sides of expressions (\ref{HC2}) and (\ref{HC3}) turn to zero, which implies that for the case of propagation of a coupled gravitational wave in empty space, the conditions of a harmonic gauge are not violated.

\section{The energy-momentum tensor in the presence of a constant magnetic field}\label{sect5}

Let a plane harmonic electromagnetic wave be emitted from a point $x=0$ in the direction of the axis $x^{1}=x$  in a medium with dielectric constant $\varepsilon$ and magnetic one $\mu=1$, which propagates in a constant magnetic field with induction $H_{00}$ having a direction along the axis $x^{3}=z$. Then expressions (\ref{EF}) and (\ref{MF}) for the electric and magnetic field intensities take the form
\begin{eqnarray}
\label{EFH}
&&E_{y}=E_{0}\cos\left(\omega\left(t-\frac{nx}{c}\right)\right),\\
\label{MFH}
&&H_{z}=H_{0}\cos\left(\omega\left(t-\frac{nx}{c}\right)\right)+H_{00}.
\end{eqnarray}

The non-zero components of the energy-momentum tensor $T_{\mu\nu}$ for the case under consideration have the following form
\begin{eqnarray}
\label{T00SH}
&&{T_{00}} = \frac{1}{{8\pi }}\left( {\varepsilon E_y^2 + H_z^2} \right) = \frac{{{n^2}E_0^2}}{{8\pi }}\left( {1 + \cos \left( {2\omega \left( {t - \frac{{nx}}{c}} \right)} \right)} \right) + \frac{1}{{8\pi }}\left( {H_{00}^2 + 2n{E_0}{H_{00}}\cos \left( {\omega \left( {t - \frac{{nx}}{c}} \right)} \right)} \right),\\
\label{T01SH}
&&{T_{01}} = {T_{10}} =  - \frac{1}{{4\pi }}\varepsilon {E_y}{H_z} - \frac{1}{{4\pi }}{E_y}{H_{00}} =  - \frac{{nE_0^2}}{{8\pi }}\left( {1 + \cos \left( {2\omega \left( {t - \frac{{nx}}{c}} \right)} \right)} \right) - \frac{{{E_0}{H_{00}}}}{{4\pi }}\cos \left( {\omega \left( {t - \frac{{nx}}{c}} \right)} \right),\\
\label{T11SH}
&&{T_{11}} = \frac{1}{{8\pi }}\left( {E_y^2 + \frac{{H_z^2}}{\varepsilon }} \right) = \frac{{E_0^2}}{{8\pi }}\left( {1 + \cos \left( {2\omega \left( {t - \frac{{nx}}{c}} \right)} \right)} \right) + \frac{1}{{8\pi }}\left( {\frac{{H_{00}^2}}{{{n^2}}} + \frac{{2{E_0}{H_{00}}}}{n}\cos \left( {\omega \left( {t - \frac{{nx}}{c}} \right)} \right)} \right),\\
\label{T22SH}
\nonumber
&&{T_{22}} = \frac{1}{{8\pi }}\left( {E_y^2 + \frac{{H_z^2}}{\varepsilon }} \right) - \frac{1}{{4\pi }}\varepsilon E_y^2 = \frac{{\left( {1 - {n^2}} \right)E_0^2}}{{8\pi }}\left( {1 + \cos \left( {2\omega \left( {t - \frac{{nx}}{c}} \right)} \right)} \right) \\
&&+ \frac{1}{{8\pi }}\left( {\frac{{H_{00}^2}}{{{n^2}}} + \frac{{2{E_0}{H_{00}}}}{n}\cos \left( {\omega \left( {t - \frac{{nx}}{c}} \right)} \right)} \right),\\
\label{T33SH}
\nonumber
&&{T_{33}} = \frac{1}{{8\pi }}\left( {E_y^2 + \frac{{H_z^2}}{\varepsilon }} \right) - \frac{1}{{4\pi }}H_z^2 = \frac{{\left( {1 - {n^2}} \right)E_0^2}}{{8\pi }}\left( {1 + \cos \left( {2\omega \left( {t - \frac{{nx}}{c}} \right)} \right)} \right) \\
&&+ \frac{1}{{8\pi }}\left( {\frac{{1 - 2{n^2}}}{{{n^2}}}H_{00}^2 + \frac{{2\left( {1 - 2{n^2}} \right){E_0}{H_{00}}}}{n}\cos \left( {\omega \left( {t - \frac{{nx}}{c}} \right)} \right)} \right).
\end{eqnarray}

We also note that the energy-momentum tensor with components (\ref{T00SH})--(\ref{T33SH}) satisfies the conservation laws (\ref{CL}).

Formulas (\ref{T00SH}), (\ref{T11SH})--(\ref{T33SH}), in this case, gives the following expression for the scalar $T$ in the case of a small curvature of space-time
\begin{eqnarray}
\label{ScalarTH}
\nonumber
T=\eta_{\mu\nu}T^{\mu\nu}=T_{00}-T_{11}-T_{22}-T_{33}=
\frac{{3\left( {{n^2} - 1} \right)E_0^2}}{{8\pi }}\left( {1 + \cos \left( {2\omega \left( {t - \frac{{nx}}{c}} \right)} \right)} \right)\\
 + \frac{1}{{8\pi }}\left( {\frac{{3\left( {{n^2} - 1} \right)}}{{{n^2}}}H_{00}^2 + \frac{{6\left( {{n^2} - 1} \right)}}{n}{E_0}{H_{00}}\cos \left( {\omega \left( {t - \frac{{nx}}{c}} \right)} \right)} \right).
\end{eqnarray}

Then, using expressions (\ref{TC}), (\ref{T00SH}), (\ref{T11SH})--(\ref{ScalarTH}) one has
\begin{eqnarray}
\label{T00H}
&&{\tilde T_{00}} = \frac{{\left( {3 - {n^2}} \right)E_0^2}}{{16\pi }}\left( {1 + \cos \left( {2\omega \left( {t - \frac{{nx}}{c}} \right)} \right)} \right) + \frac{1}{{8\pi }}\left( {\frac{{3 - {n^2}}}{{2{n^2}}}H_{00}^2 + \frac{{3 - {n^2}}}{n}{E_0}{H_{00}}\cos \left( {\omega \left( {t - \frac{{nx}}{c}} \right)} \right)} \right),\\
\label{T11H}
&&{\tilde T_{11}} = \frac{{\left( {3{n^2} - 1} \right)E_0^2}}{{16\pi }}\left( {1 + \cos \left( {2\omega \left( {t - \frac{{nx}}{c}} \right)} \right)} \right) + \frac{1}{{8\pi }}\left( {\frac{{5{n^2} - 3}}{{2{n^2}}}H_{00}^2 + \frac{{3{n^2} - 1}}{n}{E_0}{H_{00}}\cos \left( {\omega \left( {t - \frac{{nx}}{c}} \right)} \right)} \right),\\
\label{T22H}
&&{\tilde T_{22}} = \frac{{\left( {{n^2} - 1} \right)E_0^2}}{{16\pi }}\left( {1 + \cos \left( {2\omega \left( {t - \frac{{nx}}{c}} \right)} \right)} \right) + \frac{1}{{8\pi }}\left( {\frac{{5{n^2} - 3}}{{2{n^2}}}H_{00}^2 + \frac{{3{n^2} - 1}}{n}{E_0}{H_{00}}\cos \left( {\omega \left( {t - \frac{{nx}}{c}} \right)} \right)} \right),\\
\label{T33H}
&&{\tilde T_{33}} = \frac{{\left( {{n^2} - 1} \right)E_0^2}}{{16\pi }}\left( {1 + \cos \left( {2\omega \left( {t - \frac{{nx}}{c}} \right)} \right)} \right) + \frac{1}{{8\pi }}\left( { - \frac{{{n^2} + 1}}{{2{n^2}}}H_{00}^2 - \frac{{{n^2} + 1}}{n}{E_0}{H_{00}}\cos \left( {\omega \left( {t - \frac{{nx}}{c}} \right)} \right)} \right),
\end{eqnarray}
and, taking into account the expression (\ref{T01SH}), one has $\tilde{T}_{01}=\tilde{T}_{10}=T_{01}=T_{10}$.

Thus, after substituting the components of the momentum energy tensor (\ref{T00H})--(\ref{T33H}) into the Einstein-Maxwell equations (\ref{E1})--(\ref{E6}), one can find gravitational-wave solutions using an approach similar to that considered in Sec. \ref{sect4}.

\section{Bound gravitational waves in a dielectric medium and a constant magnetic field}\label{sect6}

Formulas (\ref{E1})--(\ref{E6}) allow one to write down the system of Einstein-Maxwell equations for finding the components $h_{22}$, $h_{33}$ and the function $\tilde{h}$ for the propagation of an electromagnetic wave in a dielectric medium in a constant magnetic field on the basis of equations
\begin{eqnarray}
\label{E9H}
&&\frac{{{\partial ^2}\tilde h}}{{c\partial t\partial x}} =  - \frac{{2nGE_0^2}}{{{c^4}}}\left( {1 + \cos \left( {2\omega \left( {t - \frac{{nx}}{c}} \right)} \right)} \right) - \frac{{4G{E_0}{H_{00}}}}{{{c^4}}}\cos \left( {\omega \left( {t - \frac{{nx}}{c}} \right)} \right),\\
\label{E10H}
\nonumber
&&\frac{{{\partial ^2}\tilde h}}{{\partial {x^2}}} + \frac{{{\partial ^2}\tilde h}}{{{c^2}\partial {t^2}}} = \frac{{2\left( {{n^2} + 1} \right)GE_0^2}}{{{c^4}}}\left( {1 + \cos \left( {2\omega \left( {t - \frac{{nx}}{c}} \right)} \right)} \right)\\
&&+ \frac{{4G}}{{{c^4}}}\left( {H_{00}^2 + \frac{{\left( {{n^2} + 1} \right){E_0}{H_{00}}}}{n}\cos \left( {\omega \left( {t - \frac{{nx}}{c}} \right)} \right)} \right),\\
\label{E11H}
\nonumber
&&\frac{{{\partial ^2}{h_{22}}}}{{\partial {x^2}}} - \frac{{{\partial ^2}{h_{22}}}}{{{c^2}\partial {t^2}}} = \frac{{\left( {{n^2} - 1} \right)GE_0^2}}{{{c^4}}}\left( {1 + \cos \left( {2\omega \left( {t - \frac{{nx}}{c}} \right)} \right)} \right)\\
&& + \frac{{2G}}{{{c^4}}}\left( {\frac{{5{n^2} - 3}}{{2{n^2}}}H_{00}^2 + \frac{{\left( {3{n^2} - 1} \right){E_0}{H_{00}}}}{n}\cos \left( {\omega \left( {t - \frac{{nx}}{c}} \right)} \right)}\right),\\
\label{E12H}
\nonumber
&&\frac{{{\partial ^2}{h_{33}}}}{{\partial {x^2}}} - \frac{{{\partial ^2}{h_{33}}}}{{{c^2}\partial {t^2}}} = \frac{{\left( {{n^2} - 1} \right)GE_0^2}}{{{c^4}}}\left( {1 + \cos \left( {2\omega \left( {t - \frac{{nx}}{c}} \right)} \right)} \right) \\
&&+ \frac{{2G}}{{{c^4}}}\left( { - \frac{{{n^2} + 1}}{{2{n^2}}}H_{00}^2 - \frac{{\left( {{n^2} + 1} \right){E_0}{H_{00}}}}{n}\cos \left( {\omega \left( {t - \frac{{nx}}{c}} \right)} \right)} \right).
\end{eqnarray}

The solution of equations (\ref{E9H})--(\ref{E12H}) for the case of describing only gravitational waves at a frequency $\omega$ gives
\begin{eqnarray}
\label{S31H}
&&{h_{22}} =  - \frac{{4nG{E_0}{H_{00}}}}{{\left( {{n^2} - 1} \right){c^2}{\omega ^2}}}\left( {\cos \left( {\omega \left( {t - \frac{{nx}}{c}} \right)} \right) - \cos \left( {\omega \left( {t - \frac{x}{c}} \right)} \right)} \right) - \frac{{2G{E_0}{H_{00}}}}{{n{c^2}{\omega ^2}}}\cos \left( {\omega \left( {t - \frac{{nx}}{c}} \right)} \right),\\
\label{S41H}
&&{h_{33}} = \frac{{4nG{E_0}{H_{00}}}}{{\left( {{n^2} - 1} \right){c^2}{\omega ^2}}}\left( {\cos \left( {\omega \left( {t - \frac{{nx}}{c}} \right)} \right) - \cos \left( {\omega \left( {t - \frac{x}{c}} \right)} \right)} \right) - \frac{{2G{E_0}{H_{00}}}}{{n{c^2}{\omega ^2}}}\cos \left( {\omega \left( {t - \frac{{nx}}{c}} \right)} \right),\\
\label{S51H}
&&\tilde h = {h_{22}} + {h_{33}} =  - \frac{{4G{E_0}{H_{00}}}}{{n{c^2}{\omega ^2}}}\cos \left( {\omega \left( {t - \frac{{nx}}{c}} \right)} \right).
\end{eqnarray}

The first terms in formulas (\ref{S31H}) and (\ref{S41H}) are equal to each other, but have the opposite sign. They describe a transverse gravitational wave in a dielectric medium. From these expressions it follows that during the propagation of an electromagnetic wave in a dielectric and in a constant magnetic field, gravitational waves are excited with propagation velocities equal to both the speed of light in a dielectric $v_{g}=v_{e}=c/n$ and the speed of light in vacuum $v_{g}=c$.

Substituting solution (\ref{S51H}) into equation (\ref{E2}) taking into account (\ref{T11H}) allows one to write the equations for the components $h_{00}$, $h_{01}=h_{10}$ and $h_{11}$, taking into account only the wave components of the coupled gravitational wave in the following form
\begin{equation}
\label{EM1H}
\frac{{{\partial ^2}{h_{11}}}}{{{c^2}\partial {t^2}}} - 2\frac{{{\partial ^2}{h_{01}}}}{{c\partial t\partial x}} + \frac{{{\partial ^2}{h_{00}}}}{{\partial {x^2}}} = \frac{{2\left( {{n^2} - 1} \right)G{E_0}{H_{00}}}}{{n{c^4}}}\cos \left( {\omega \left( {t - \frac{{nx}}{c}} \right)} \right).
\end{equation}
The solution to this equation has the form
\begin{eqnarray}
\label{S3H}
&&{h_{00}} =  - \frac{{2G{E_0}{H_{00}}}}{{n{c^3}\omega }}x\sin \left( {\omega \left( {t - \frac{{nx}}{c}} \right)} \right) + \frac{{{\rm{2}}G{E_0}{H_{00}}}}{{n{c^2}{\omega ^2}}}\cos \left( {\omega \left( {t - \frac{{nx}}{c}} \right)} \right),\\
\label{S4H}
&&{h_{01}} = {h_{10}} = \frac{{{\rm{2}}G{E_0}{H_{00}}}}{{{c^3}\omega }}x\sin \left( {\omega \left( {t - \frac{{nx}}{c}} \right)} \right) - \frac{{{\rm{2}}G{E_0}{H_{00}}}}{{{c^2}{\omega ^2}}}\cos \left( {\omega \left( {t - \frac{{nx}}{c}} \right)} \right),\\
\label{S5H}
&&{h_{11}} =  - \frac{{2nG{E_0}{H_{00}}}}{{{c^3}\omega }}x\sin \left( {\omega \left( {t - \frac{{nx}}{c}} \right)} \right) + \frac{{{\rm{2}}G{E_0}{H_{00}}}}{{n{c^2}{\omega ^2}}}\cos \left( {\omega \left( {t - \frac{{nx}}{c}} \right)} \right).
\end{eqnarray}

It follows from the obtained solutions (\ref{S31H}), (\ref{S41H}) and (\ref{S3H})--(\ref{S5H}) that the speed of a coupled gravitational wave in a dielectric medium propagating in a constant magnetic field is less than the speed of light in empty space and is equal to the speed of an electromagnetic wave $v_{g}=v_{e}=c/n$.

Substitution of expressions (\ref{S31H}), (\ref{S41H}) and (\ref{S3H})--(\ref{S5H}) into formula (\ref{EnFlux}) under the condition $x\gg\frac{\omega}{c}$ allows, as a first approximation, to obtain the expression for the energy flux density of the gravitational wave
\begin{eqnarray}
\label{FDH}
\nonumber
&&c{t^{01}} = \frac{{{{\left( {{n^2} - 1} \right)}^2}GE_0^2H_{00}^2}}{{8\pi n{c^3}}}{x^2}{\cos ^2}\left( {\omega \left( {t - \frac{{nx}}{c}} \right)} \right) \\
&&+ \frac{{{n^2}GE_0^2H_{00}^2}}{{\pi {{\left( {{n^2} - 1} \right)}^2}c{\omega ^2}}}\left( {\sin \left( {\omega \left( {t - \frac{{nx}}{c}} \right)} \right) - \sin \left( {\omega \left( {t - \frac{x}{c}} \right)} \right)} \right)\left( {n\sin \left( {\omega \left( {t - \frac{{nx}}{c}} \right)} \right) - \sin \left( {\omega \left( {t - \frac{x}{c}} \right)} \right)} \right).
\end{eqnarray}

We also note, that when $n\rightarrow1$, the energy flux density of the gravitational wave equal to
\begin{equation}
\label{FDH1}
c{t^{01}} = \frac{{GE_0^2H_{00}^2}}{{4\pi {c^3}}}{x^2}{\cos ^2}\left( {\omega \left( {t - \frac{x}{c}} \right)} \right).
\end{equation}

As one can see, solutions (\ref{S31H}), (\ref{S41H}) and (\ref{S3H})--(\ref{S5H}) don't satisfy the harmonic gauge (\ref{HC}), namely
\begin{eqnarray}
\label{HC2H}
&&\frac{\partial }{{c\partial t}}\left( {{h_{00}} - \frac{1}{2}h} \right) - \frac{{\partial {h_{10}}}}{{\partial x}} = \frac{{\left( {{n^2} - 1} \right)G{E_0}{H_{00}}}}{{n{c^4}}}x\cos \left( {\omega \left( {t - \frac{{nx}}{c}} \right)} \right) + \frac{{2\left( {n - 1} \right)G{E_0}{H_{00}}}}{{{c^3}\omega }}\sin \left( {\omega \left( {t - \frac{{nx}}{c}} \right)} \right),\\
\label{HC3H}
&&\frac{{\partial {h_{01}}}}{{c\partial t}} - \frac{\partial }{{\partial x}}\left( {{h_{11}} + \frac{1}{2}h} \right) = \frac{{\left( {1 - {n^2}} \right)G{E_0}{H_{00}}}}{{{c^4}}}x\cos \left( {\omega \left( {t - \frac{{nx}}{c}} \right)} \right) + \frac{{2{{\left( {n - 1} \right)}^2}G{E_0}{H_{00}}}}{{{c^3}\omega }}\sin \left( {\omega \left( {t - \frac{{nx}}{c}} \right)} \right).
\end{eqnarray}
where
\begin{equation}
\label{HCAH}
h=h_{00}-h_{11}-h_{22}-h_{33}=
\frac{{2\left( {{n^2} - 1} \right)G{E_0}{H_{00}}}}{{n{c^3}\omega }}x\sin \left( {\omega \left( {t - \frac{{n\omega }}{c}} \right)} \right) + \frac{{4G{E_0}{H_{00}}}}{{n{c^2}{\omega ^2}}}\cos \left( {\omega \left( {t - \frac{{n\omega }}{c}} \right)} \right).
\end{equation}

Since the right-hand sides of these expressions are not equal to zero, harmonic gauge (\ref{HC}) for solutions  (\ref{S31H}), (\ref{S41H}) and (\ref{S3H})--(\ref{S5H}) is violated. However, for the case $n\rightarrow1$, the right-hand sides of expressions (\ref{HC2H}) and (\ref{HC3H}) turn to zero, which implies that for the case of propagation of a coupled gravitational wave in empty space, the conditions of a harmonic gauge are satisfied.

Expressions (\ref{S31H}) and (\ref{S41H}) allow one to consider the limiting case when $n\rightarrow1$ corresponding to empty space where the components $h_{22}=-h_{33}$ \cite{Landau,Weber, Maggiore:2018sht}. Thus, from the first terms of formulas (\ref{S31H}) and (\ref{S41H}) after calculating the uncertainty we obtain
\begin{equation}
\label{HCAHL}
{h_{22}} =  - {h_{33}} = {\left. { - \frac{{4nG{E_0}{H_{00}}}}{{\left( {n + 1} \right){c^2}{\omega ^2}}}\frac{{\cos \left( {\omega \left( {t - \frac{{nx}}{c}} \right)} \right) - \cos \left( {\omega \left( {t - \frac{x}{c}} \right)} \right)}}{{n - 1}}} \right|_{n \to 1}} =  - \frac{{2G{E_0}{H_{00}}}}{{{c^3}\omega }}x\sin \left( {\omega \left( {t - \frac{x}{c}} \right)} \right).
\end{equation}

Formula (\ref{HCAHL}) coincides with the well-known expression obtained when calculating the propagation of an electromagnetic wave in a constant magnetic field in empty space \cite{Kolosnitsyn:2015zua}.
Therefore, the obtained  solutions for the transverse gravitational-wave components are a generalization of previously known ones to the case of dielectric media with a permittivity (\ref{vareps}).
We also note that the substitution of formula  (\ref{HCAHL}) into (\ref{EnFlux}) for $x\gg\frac{\omega}{c}$ leads to expression (\ref{FDH1}).

\section{Conclusion}

The description of the propagation of a bound longitudinal transverse gravitational wave excited by a strong electromagnetic wave propagating in a dielectric medium both without a constant magnetic field and in the presence of this field carried out in this work showed that the speed of such a gravitational wave differs from the speed of light in an empty space and is equal to the speed of the electromagnetic wave in a medium with a dielectric $v_{g}=v_{e}=c/n$. In the presence of a constant magnetic field, transverse gravitational waves propagating with the speed of light in a vacuum $v_{g}=c$ are excited as well.

The bound gravitational waves in a  dielectric medium can be attributed to the class of forced waves that exist due to propagation of forcing electromagnetic waves. Such a waves cannot exist in a dielectric medium without the presence of electromagnetic waves generating them.

It should be noted that the formulas obtained to describe the propagation of an electromagnetic wave in a dielectric medium without a constant magnetic field for $n\rightarrow1$ go over into the expressions given in \cite{Morozov2020} for the case of a gravitational wave coupled with electromagnetic one in empty space.

\begin{acknowledgments}
The study was funded by a grant from the Russian Science Foundation (project No. 19-12-00242).
\end{acknowledgments}

\bibliography{ref}

\begin{thebibliography}{39}%
\makeatletter
\providecommand \@ifxundefined [1]{%
 \@ifx{#1\undefined}
}%
\providecommand \@ifnum [1]{%
 \ifnum #1\expandafter \@firstoftwo
 \else \expandafter \@secondoftwo
 \fi
}%
\providecommand \@ifx [1]{%
 \ifx #1\expandafter \@firstoftwo
 \else \expandafter \@secondoftwo
 \fi
}%
\providecommand \natexlab [1]{#1}%
\providecommand \enquote  [1]{``#1''}%
\providecommand \bibnamefont  [1]{#1}%
\providecommand \bibfnamefont [1]{#1}%
\providecommand \citenamefont [1]{#1}%
\providecommand \href@noop [0]{\@secondoftwo}%
\providecommand \href [0]{\begingroup \@sanitize@url \@href}%
\providecommand \@href[1]{\@@startlink{#1}\@@href}%
\providecommand \@@href[1]{\endgroup#1\@@endlink}%
\providecommand \@sanitize@url [0]{\catcode `\\12\catcode `\$12\catcode
  `\&12\catcode `\#12\catcode `\^12\catcode `\_12\catcode `\%12\relax}%
\providecommand \@@startlink[1]{}%
\providecommand \@@endlink[0]{}%
\providecommand \url  [0]{\begingroup\@sanitize@url \@url }%
\providecommand \@url [1]{\endgroup\@href {#1}{\urlprefix }}%
\providecommand \urlprefix  [0]{URL }%
\providecommand \Eprint [0]{\href }%
\providecommand \doibase [0]{http://dx.doi.org/}%
\providecommand \selectlanguage [0]{\@gobble}%
\providecommand \bibinfo  [0]{\@secondoftwo}%
\providecommand \bibfield  [0]{\@secondoftwo}%
\providecommand \translation [1]{[#1]}%
\providecommand \BibitemOpen [0]{}%
\providecommand \bibitemStop [0]{}%
\providecommand \bibitemNoStop [0]{.\EOS\space}%
\providecommand \EOS [0]{\spacefactor3000\relax}%
\providecommand \BibitemShut  [1]{\csname bibitem#1\endcsname}%
\let\auto@bib@innerbib\@empty
\bibitem [{\citenamefont {Tolman}\ \emph {et~al.}(1931)\citenamefont {Tolman},
  \citenamefont {Ehrenfest},\ and\ \citenamefont {Podolsky}}]{Tolman:1931zza}%
  \BibitemOpen
  \bibfield  {author} {\bibinfo {author} {\bibfnamefont {R.~C.}\ \bibnamefont
  {Tolman}}, \bibinfo {author} {\bibfnamefont {P.}~\bibnamefont {Ehrenfest}}, \
  and\ \bibinfo {author} {\bibfnamefont {B.}~\bibnamefont {Podolsky}},\ }\href
  {\doibase 10.1103/PhysRev.37.602} {\bibfield  {journal} {\bibinfo  {journal}
  {Phys. Rev.}\ }\textbf {\bibinfo {volume} {37}},\ \bibinfo {pages} {602}
  (\bibinfo {year} {1931})}\BibitemShut {NoStop}%
\bibitem [{\citenamefont {Hegarty}(1969)}]{Hegarty1969}%
  \BibitemOpen
  \bibfield  {author} {\bibinfo {author} {\bibfnamefont {J.~C.}\ \bibnamefont
  {Hegarty}},\ }\href {\doibase 10.1007/BF02711695} {\bibfield  {journal}
  {\bibinfo  {journal} {Nuovo Cimento B}\ }\textbf {\bibinfo {volume} {61}},\
  \bibinfo {pages} {47} (\bibinfo {year} {1969})}\BibitemShut {NoStop}%
\bibitem [{\citenamefont {Bonnor}(1969)}]{Bonnor1969}%
  \BibitemOpen
  \bibfield  {author} {\bibinfo {author} {\bibfnamefont {W.~B.}\ \bibnamefont
  {Bonnor}},\ }\href {\doibase 10.1007/BF01645484} {\bibfield  {journal}
  {\bibinfo  {journal} {Commun.Math. Phys.}\ }\textbf {\bibinfo {volume}
  {13}},\ \bibinfo {pages} {163} (\bibinfo {year} {1969})}\BibitemShut
  {NoStop}%
\bibitem [{\citenamefont {Aichelburg}\ and\ \citenamefont
  {Sexl}(1971)}]{Aichelburg:1970dh}%
  \BibitemOpen
  \bibfield  {author} {\bibinfo {author} {\bibfnamefont {P.~C.}\ \bibnamefont
  {Aichelburg}}\ and\ \bibinfo {author} {\bibfnamefont {R.~U.}\ \bibnamefont
  {Sexl}},\ }\href {\doibase 10.1007/BF00758149} {\bibfield  {journal}
  {\bibinfo  {journal} {Gen. Rel. Grav.}\ }\textbf {\bibinfo {volume} {2}},\
  \bibinfo {pages} {303} (\bibinfo {year} {1971})}\BibitemShut {NoStop}%
\bibitem [{\citenamefont {Nackoney}(1973)}]{Nackoney1973}%
  \BibitemOpen
  \bibfield  {author} {\bibinfo {author} {\bibfnamefont {R.~W.}\ \bibnamefont
  {Nackoney}},\ }\href {\doibase 10.1063/1.1666473} {\bibfield  {journal}
  {\bibinfo  {journal} {Journal of Mathematical Physics}\ }\textbf {\bibinfo
  {volume} {14}},\ \bibinfo {pages} {1239} (\bibinfo {year}
  {1973})}\BibitemShut {NoStop}%
\bibitem [{\citenamefont {Voronov}\ and\ \citenamefont
  {Kobzarev}(1974)}]{Voronov:1974}%
  \BibitemOpen
  \bibfield  {author} {\bibinfo {author} {\bibfnamefont {N.~A.}\ \bibnamefont
  {Voronov}}\ and\ \bibinfo {author} {\bibfnamefont {I.~Y.}\ \bibnamefont
  {Kobzarev}},\ }\href@noop {} {\bibfield  {journal} {\bibinfo  {journal} {Zh.
  Eksp. Teor. Fiz.}\ }\textbf {\bibinfo {volume} {68}},\ \bibinfo {pages}
  {1179} (\bibinfo {year} {1974})}\BibitemShut {NoStop}%
\bibitem [{\citenamefont {Scully}(1979)}]{Scully:1979xe}%
  \BibitemOpen
  \bibfield  {author} {\bibinfo {author} {\bibfnamefont {M.~O.}\ \bibnamefont
  {Scully}},\ }\href {\doibase 10.1103/PhysRevD.19.3582} {\bibfield  {journal}
  {\bibinfo  {journal} {Phys. Rev.}\ }\textbf {\bibinfo {volume} {D19}},\
  \bibinfo {pages} {3582} (\bibinfo {year} {1979})}\BibitemShut {NoStop}%
\bibitem [{\citenamefont {Dray}\ and\ \citenamefont
  {'t~Hooft}(1985)}]{Dray:1984ha}%
  \BibitemOpen
  \bibfield  {author} {\bibinfo {author} {\bibfnamefont {T.}~\bibnamefont
  {Dray}}\ and\ \bibinfo {author} {\bibfnamefont {G.}~\bibnamefont
  {'t~Hooft}},\ }\href {\doibase 10.1016/0550-3213(85)90525-5} {\bibfield
  {journal} {\bibinfo  {journal} {Nucl. Phys.}\ }\textbf {\bibinfo {volume}
  {B253}},\ \bibinfo {pages} {173} (\bibinfo {year} {1985})}\BibitemShut
  {NoStop}%
\bibitem [{\citenamefont {Bonnor}(2009)}]{Bonnor:2009zza}%
  \BibitemOpen
  \bibfield  {author} {\bibinfo {author} {\bibfnamefont {W.~B.}\ \bibnamefont
  {Bonnor}},\ }\href {\doibase 10.1007/s10714-008-0655-z} {\bibfield  {journal}
  {\bibinfo  {journal} {Gen. Rel. Grav.}\ }\textbf {\bibinfo {volume} {41}},\
  \bibinfo {pages} {77} (\bibinfo {year} {2009})}\BibitemShut {NoStop}%
\bibitem [{\citenamefont {van Holten}(2011)}]{vanHolten:2008ts}%
  \BibitemOpen
  \bibfield  {author} {\bibinfo {author} {\bibfnamefont {J.~W.}\ \bibnamefont
  {van Holten}},\ }\href {\doibase 10.1002/prop.201000088} {\bibfield
  {journal} {\bibinfo  {journal} {Fortsch. Phys.}\ }\textbf {\bibinfo {volume}
  {59}},\ \bibinfo {pages} {284} (\bibinfo {year} {2011})},\ \Eprint
  {http://arxiv.org/abs/0808.0997} {arXiv:0808.0997 [hep-ph]} \BibitemShut
  {NoStop}%
\bibitem [{\citenamefont {Ratzel}\ \emph {et~al.}(2016)\citenamefont {Ratzel},
  \citenamefont {Wilkens},\ and\ \citenamefont {Menzel}}]{Ratzel:2015nqf}%
  \BibitemOpen
  \bibfield  {author} {\bibinfo {author} {\bibfnamefont {D.}~\bibnamefont
  {Ratzel}}, \bibinfo {author} {\bibfnamefont {M.}~\bibnamefont {Wilkens}}, \
  and\ \bibinfo {author} {\bibfnamefont {R.}~\bibnamefont {Menzel}},\ }\href
  {\doibase 10.1088/1367-2630/18/2/023009} {\bibfield  {journal} {\bibinfo
  {journal} {New J. Phys.}\ }\textbf {\bibinfo {volume} {18}},\ \bibinfo
  {pages} {023009} (\bibinfo {year} {2016})},\ \Eprint
  {http://arxiv.org/abs/1511.01023} {arXiv:1511.01023 [quant-ph]} \BibitemShut
  {NoStop}%
\bibitem [{\citenamefont {van Holten}(2018)}]{vanHolten:2018izl}%
  \BibitemOpen
  \bibfield  {author} {\bibinfo {author} {\bibfnamefont {J.~W.}\ \bibnamefont
  {van Holten}},\ }\href {\doibase 10.3390/universe4100110} {\bibfield
  {journal} {\bibinfo  {journal} {Universe}\ }\textbf {\bibinfo {volume} {4}},\
  \bibinfo {pages} {110} (\bibinfo {year} {2018})},\ \Eprint
  {http://arxiv.org/abs/1809.04309} {arXiv:1809.04309 [gr-qc]} \BibitemShut
  {NoStop}%
\bibitem [{\citenamefont {Eddington}(1922)}]{Eddington1}%
  \BibitemOpen
  \bibfield  {author} {\bibinfo {author} {\bibfnamefont {A.~S.}\ \bibnamefont
  {Eddington}},\ }\href@noop {} {\bibfield  {journal} {\bibinfo  {journal}
  {Proc. Roy. Soc. London A}\ }\textbf {\bibinfo {volume} {102}},\ \bibinfo
  {pages} {268} (\bibinfo {year} {1922})}\BibitemShut {NoStop}%
\bibitem [{\citenamefont {Gertsenshtein}(1962)}]{Gertsenshtein1962}%
  \BibitemOpen
  \bibfield  {author} {\bibinfo {author} {\bibfnamefont {M.~E.}\ \bibnamefont
  {Gertsenshtein}},\ }\href@noop {} {\bibfield  {journal} {\bibinfo  {journal}
  {Sov. Phys. JETP}\ }\textbf {\bibinfo {volume} {14}},\ \bibinfo {pages} {84}
  (\bibinfo {year} {1962})}\BibitemShut {NoStop}%
\bibitem [{\citenamefont {Grishchuk}\ and\ \citenamefont
  {Sazhin}(1973)}]{Grishchuk:1973qz}%
  \BibitemOpen
  \bibfield  {author} {\bibinfo {author} {\bibfnamefont {L.~P.}\ \bibnamefont
  {Grishchuk}}\ and\ \bibinfo {author} {\bibfnamefont {M.~V.}\ \bibnamefont
  {Sazhin}},\ }\href@noop {} {\bibfield  {journal} {\bibinfo  {journal} {Zh.
  Eksp. Teor. Fiz.}\ }\textbf {\bibinfo {volume} {65}},\ \bibinfo {pages} {441}
  (\bibinfo {year} {1973})}\BibitemShut {NoStop}%
\bibitem [{\citenamefont {Grishchuk}\ and\ \citenamefont
  {Sazhin}(1975)}]{Grishchuk:1975}%
  \BibitemOpen
  \bibfield  {author} {\bibinfo {author} {\bibfnamefont {L.~P.}\ \bibnamefont
  {Grishchuk}}\ and\ \bibinfo {author} {\bibfnamefont {M.~V.}\ \bibnamefont
  {Sazhin}},\ }\href@noop {} {\bibfield  {journal} {\bibinfo  {journal} {Zh.
  Eksp. Teor. Fiz.}\ }\textbf {\bibinfo {volume} {68}},\ \bibinfo {pages}
  {1569} (\bibinfo {year} {1975})}\BibitemShut {NoStop}%
\bibitem [{\citenamefont {Pustovoit}\ and\ \citenamefont
  {Chernozatonsky}(1981)}]{Pustovoit:1981za}%
  \BibitemOpen
  \bibfield  {author} {\bibinfo {author} {\bibfnamefont {V.~I.}\ \bibnamefont
  {Pustovoit}}\ and\ \bibinfo {author} {\bibfnamefont {L.~A.}\ \bibnamefont
  {Chernozatonsky}},\ }\href@noop {} {\bibfield  {journal} {\bibinfo  {journal}
  {Zh. Eksp. Teor. Fiz.}\ }\textbf {\bibinfo {volume} {34}},\ \bibinfo {pages}
  {241} (\bibinfo {year} {1981})}\BibitemShut {NoStop}%
\bibitem [{\citenamefont {Nikishov}\ and\ \citenamefont
  {Ritus}(2011)}]{Nikishov:2010zz}%
  \BibitemOpen
  \bibfield  {author} {\bibinfo {author} {\bibfnamefont {A.~I.}\ \bibnamefont
  {Nikishov}}\ and\ \bibinfo {author} {\bibfnamefont {V.~I.}\ \bibnamefont
  {Ritus}},\ }\href {\doibase 10.3367/UFNe.0180.201011b.1135} {\bibfield
  {journal} {\bibinfo  {journal} {Phys. Usp.}\ }\textbf {\bibinfo {volume}
  {53}},\ \bibinfo {pages} {1093} (\bibinfo {year} {2011})},\ \bibinfo {note}
  {[Usp. Fiz. Nauk180,1135(2010)]}\BibitemShut {NoStop}%
\bibitem [{\citenamefont {Morozov}\ and\ \citenamefont
  {Pustovoit}(2020)}]{Morozov2020}%
  \BibitemOpen
  \bibfield  {author} {\bibinfo {author} {\bibfnamefont {A.~N.}\ \bibnamefont
  {Morozov}}\ and\ \bibinfo {author} {\bibfnamefont {V.~I.}\ \bibnamefont
  {Pustovoit}},\ }\href@noop {} {\bibfield  {journal} {\bibinfo  {journal}
  {Herald of the Bauman Moscow State Technical University, Series Natural
  Sciences}\ }\textbf {\bibinfo {volume} {1(88)}},\ \bibinfo {pages} {46}
  (\bibinfo {year} {2020})}\BibitemShut {NoStop}%
\bibitem [{\citenamefont {Morozov}\ \emph {et~al.}(2020)\citenamefont
  {Morozov}, \citenamefont {Pustovoit},\ and\ \citenamefont
  {Fomin}}]{Morozov:2020snv}%
  \BibitemOpen
  \bibfield  {author} {\bibinfo {author} {\bibfnamefont {A.~N.}\ \bibnamefont
  {Morozov}}, \bibinfo {author} {\bibfnamefont {V.~I.}\ \bibnamefont
  {Pustovoit}}, \ and\ \bibinfo {author} {\bibfnamefont {I.~V.}\ \bibnamefont
  {Fomin}},\ }\href@noop {} {\  (\bibinfo {year} {2020})},\ \Eprint
  {http://arxiv.org/abs/2003.04104} {arXiv:2003.04104 [gr-qc]} \BibitemShut
  {NoStop}%
\bibitem [{\citenamefont {Abbott}\ \emph
  {et~al.}(2016{\natexlab{a}})\citenamefont {Abbott} \emph
  {et~al.}}]{Abbott:2016blz}%
  \BibitemOpen
  \bibfield  {author} {\bibinfo {author} {\bibfnamefont {B.~P.}\ \bibnamefont
  {Abbott}} \emph {et~al.} (\bibinfo {collaboration} {LIGO Scientific,
  Virgo}),\ }\href {\doibase 10.1103/PhysRevLett.116.061102} {\bibfield
  {journal} {\bibinfo  {journal} {Phys. Rev. Lett.}\ }\textbf {\bibinfo
  {volume} {116}},\ \bibinfo {pages} {061102} (\bibinfo {year}
  {2016}{\natexlab{a}})},\ \Eprint {http://arxiv.org/abs/1602.03837}
  {arXiv:1602.03837 [gr-qc]} \BibitemShut {NoStop}%
\bibitem [{\citenamefont {Abbott}\ \emph
  {et~al.}(2016{\natexlab{b}})\citenamefont {Abbott} \emph
  {et~al.}}]{Abbott:2016nmj}%
  \BibitemOpen
  \bibfield  {author} {\bibinfo {author} {\bibfnamefont {B.~P.}\ \bibnamefont
  {Abbott}} \emph {et~al.} (\bibinfo {collaboration} {LIGO Scientific,
  Virgo}),\ }\href {\doibase 10.1103/PhysRevLett.116.241103} {\bibfield
  {journal} {\bibinfo  {journal} {Phys. Rev. Lett.}\ }\textbf {\bibinfo
  {volume} {116}},\ \bibinfo {pages} {241103} (\bibinfo {year}
  {2016}{\natexlab{b}})},\ \Eprint {http://arxiv.org/abs/1606.04855}
  {arXiv:1606.04855 [gr-qc]} \BibitemShut {NoStop}%
\bibitem [{\citenamefont {Abbott}\ \emph {et~al.}(2018)\citenamefont {Abbott}
  \emph {et~al.}}]{Aasi:2013wya}%
  \BibitemOpen
  \bibfield  {author} {\bibinfo {author} {\bibfnamefont {B.~P.}\ \bibnamefont
  {Abbott}} \emph {et~al.} (\bibinfo {collaboration} {KAGRA, LIGO Scientific,
  VIRGO}),\ }\href {\doibase 10.1007/s41114-018-0012-9, 10.1007/lrr-2016-1}
  {\bibfield  {journal} {\bibinfo  {journal} {Living Rev. Rel.}\ }\textbf
  {\bibinfo {volume} {21}},\ \bibinfo {pages} {3} (\bibinfo {year} {2018})},\
  \Eprint {http://arxiv.org/abs/1304.0670} {arXiv:1304.0670 [gr-qc]}
  \BibitemShut {NoStop}%
\bibitem [{\citenamefont {Abbott}\ \emph {et~al.}(2017)\citenamefont {Abbott}
  \emph {et~al.}}]{Monitor:2017mdv}%
  \BibitemOpen
  \bibfield  {author} {\bibinfo {author} {\bibfnamefont {B.~P.}\ \bibnamefont
  {Abbott}} \emph {et~al.} (\bibinfo {collaboration} {LIGO Scientific, Virgo,
  Fermi-GBM, INTEGRAL}),\ }\href {\doibase 10.3847/2041-8213/aa920c} {\bibfield
   {journal} {\bibinfo  {journal} {Astrophys. J.}\ }\textbf {\bibinfo {volume}
  {848}},\ \bibinfo {pages} {L13} (\bibinfo {year} {2017})},\ \Eprint
  {http://arxiv.org/abs/1710.05834} {arXiv:1710.05834 [astro-ph.HE]}
  \BibitemShut {NoStop}%
\bibitem [{\citenamefont {Zeldovich}(1974)}]{Zel'dovich1974}%
  \BibitemOpen
  \bibfield  {author} {\bibinfo {author} {\bibfnamefont {Y.~B.}\ \bibnamefont
  {Zeldovich}},\ }\href@noop {} {\bibfield  {journal} {\bibinfo  {journal}
  {Sov. Phys. JETP}\ }\textbf {\bibinfo {volume} {38}},\ \bibinfo {pages} {652}
  (\bibinfo {year} {1974})}\BibitemShut {NoStop}%
\bibitem [{\citenamefont {Gerlach}(1974)}]{Gerlach:1974zz}%
  \BibitemOpen
  \bibfield  {author} {\bibinfo {author} {\bibfnamefont {U.~H.}\ \bibnamefont
  {Gerlach}},\ }\href {\doibase 10.1103/PhysRevLett.32.1023} {\bibfield
  {journal} {\bibinfo  {journal} {Phys. Rev. Lett.}\ }\textbf {\bibinfo
  {volume} {32}},\ \bibinfo {pages} {1023} (\bibinfo {year}
  {1974})}\BibitemShut {NoStop}%
\bibitem [{\citenamefont {Zeldovich}\ and\ \citenamefont
  {Novikov}(1983)}]{Zeldovich1983}%
  \BibitemOpen
  \bibfield  {author} {\bibinfo {author} {\bibfnamefont {Y.~B.}\ \bibnamefont
  {Zeldovich}}\ and\ \bibinfo {author} {\bibfnamefont {I.~D.}\ \bibnamefont
  {Novikov}},\ }\href@noop {} {\emph {\bibinfo {title} {{Relativistic
  Astrophysics Vol. 2}}}}\ (\bibinfo  {publisher} {University of Chicago
  Press},\ \bibinfo {address} {Chicago},\ \bibinfo {year} {1983})\BibitemShut
  {NoStop}%
\bibitem [{\citenamefont {Raffelt}\ and\ \citenamefont
  {Stodolsky}(1988)}]{Raffelt:1987im}%
  \BibitemOpen
  \bibfield  {author} {\bibinfo {author} {\bibfnamefont {G.}~\bibnamefont
  {Raffelt}}\ and\ \bibinfo {author} {\bibfnamefont {L.}~\bibnamefont
  {Stodolsky}},\ }\href {\doibase 10.1103/PhysRevD.37.1237} {\bibfield
  {journal} {\bibinfo  {journal} {Phys. Rev.}\ }\textbf {\bibinfo {volume}
  {D37}},\ \bibinfo {pages} {1237} (\bibinfo {year} {1988})}\BibitemShut
  {NoStop}%
\bibitem [{\citenamefont {Fargion}(1995)}]{Fargion:1995mm}%
  \BibitemOpen
  \bibfield  {author} {\bibinfo {author} {\bibfnamefont {D.}~\bibnamefont
  {Fargion}},\ }\href@noop {} {\bibfield  {journal} {\bibinfo  {journal} {Grav.
  Cosmol.}\ }\textbf {\bibinfo {volume} {1}},\ \bibinfo {pages} {301} (\bibinfo
  {year} {1995})},\ \Eprint {http://arxiv.org/abs/astro-ph/9604047}
  {arXiv:astro-ph/9604047 [astro-ph]} \BibitemShut {NoStop}%
\bibitem [{\citenamefont {Marklund}\ \emph {et~al.}(2000)\citenamefont
  {Marklund}, \citenamefont {Brodin},\ and\ \citenamefont
  {Dunsby}}]{Marklund:1999sp}%
  \BibitemOpen
  \bibfield  {author} {\bibinfo {author} {\bibfnamefont {M.}~\bibnamefont
  {Marklund}}, \bibinfo {author} {\bibfnamefont {G.}~\bibnamefont {Brodin}}, \
  and\ \bibinfo {author} {\bibfnamefont {P.~K.~S.}\ \bibnamefont {Dunsby}},\
  }\href {\doibase 10.1086/308957} {\bibfield  {journal} {\bibinfo  {journal}
  {Astrophys. J.}\ }\textbf {\bibinfo {volume} {536}},\ \bibinfo {pages} {875}
  (\bibinfo {year} {2000})},\ \Eprint {http://arxiv.org/abs/astro-ph/9907350}
  {arXiv:astro-ph/9907350 [astro-ph]} \BibitemShut {NoStop}%
\bibitem [{\citenamefont {Dolgov}\ and\ \citenamefont
  {Ejlli}(2012)}]{Dolgov:2012be}%
  \BibitemOpen
  \bibfield  {author} {\bibinfo {author} {\bibfnamefont {A.~D.}\ \bibnamefont
  {Dolgov}}\ and\ \bibinfo {author} {\bibfnamefont {D.}~\bibnamefont {Ejlli}},\
  }\href {\doibase 10.1088/1475-7516/2012/12/003} {\bibfield  {journal}
  {\bibinfo  {journal} {JCAP}\ }\textbf {\bibinfo {volume} {1212}},\ \bibinfo
  {pages} {003} (\bibinfo {year} {2012})},\ \Eprint
  {http://arxiv.org/abs/1211.0500} {arXiv:1211.0500 [gr-qc]} \BibitemShut
  {NoStop}%
\bibitem [{\citenamefont {Kolosnitsyn}\ and\ \citenamefont
  {Rudenko}(2015)}]{Kolosnitsyn:2015zua}%
  \BibitemOpen
  \bibfield  {author} {\bibinfo {author} {\bibfnamefont {N.~I.}\ \bibnamefont
  {Kolosnitsyn}}\ and\ \bibinfo {author} {\bibfnamefont {V.~N.}\ \bibnamefont
  {Rudenko}},\ }\href {\doibase 10.1088/0031-8949/90/7/074059} {\bibfield
  {journal} {\bibinfo  {journal} {Phys. Scripta}\ }\textbf {\bibinfo {volume}
  {90}},\ \bibinfo {pages} {074059} (\bibinfo {year} {2015})},\ \Eprint
  {http://arxiv.org/abs/1504.06548} {arXiv:1504.06548 [gr-qc]} \BibitemShut
  {NoStop}%
\bibitem [{\citenamefont {Dolgov}\ and\ \citenamefont
  {Postnov}(2017)}]{Dolgov:2017bpj}%
  \BibitemOpen
  \bibfield  {author} {\bibinfo {author} {\bibfnamefont {A.}~\bibnamefont
  {Dolgov}}\ and\ \bibinfo {author} {\bibfnamefont {K.}~\bibnamefont
  {Postnov}},\ }\href {\doibase 10.1088/1475-7516/2017/09/018} {\bibfield
  {journal} {\bibinfo  {journal} {JCAP}\ }\textbf {\bibinfo {volume} {1709}},\
  \bibinfo {pages} {018} (\bibinfo {year} {2017})},\ \Eprint
  {http://arxiv.org/abs/1706.05519} {arXiv:1706.05519 [astro-ph.HE]}
  \BibitemShut {NoStop}%
\bibitem [{\citenamefont {Ejlli}\ and\ \citenamefont
  {Thandlam}(2019)}]{Ejlli:2018hke}%
  \BibitemOpen
  \bibfield  {author} {\bibinfo {author} {\bibfnamefont {D.}~\bibnamefont
  {Ejlli}}\ and\ \bibinfo {author} {\bibfnamefont {V.~R.}\ \bibnamefont
  {Thandlam}},\ }\href {\doibase 10.1103/PhysRevD.99.044022} {\bibfield
  {journal} {\bibinfo  {journal} {Phys. Rev.}\ }\textbf {\bibinfo {volume}
  {D99}},\ \bibinfo {pages} {044022} (\bibinfo {year} {2019})},\ \Eprint
  {http://arxiv.org/abs/1807.00171} {arXiv:1807.00171 [gr-qc]} \BibitemShut
  {NoStop}%
\bibitem [{\citenamefont {Denisov}(1977)}]{Denisov1977}%
  \BibitemOpen
  \bibfield  {author} {\bibinfo {author} {\bibfnamefont {V.~I.}\ \bibnamefont
  {Denisov}},\ }\href@noop {} {\bibfield  {journal} {\bibinfo  {journal}
  {Moscow University Physics Bulletin}\ }\textbf {\bibinfo {volume} {32}},\
  \bibinfo {pages} {41} (\bibinfo {year} {1977})}\BibitemShut {NoStop}%
\bibitem [{\citenamefont {Landau}\ and\ \citenamefont
  {Lifshitz}(1987)}]{Landau}%
  \BibitemOpen
  \bibfield  {author} {\bibinfo {author} {\bibfnamefont {L.~D.}\ \bibnamefont
  {Landau}}\ and\ \bibinfo {author} {\bibfnamefont {E.~M.}\ \bibnamefont
  {Lifshitz}},\ }\href
  {https://www.elsevier.com/books/the-classical-theory-of-fields/landau/978-0-08-050349-3}
  {\emph {\bibinfo {title} {{The Classical Theory of Fields}}}}\ (\bibinfo
  {publisher} {Butterworth-Heinemann},\ \bibinfo {year} {1987})\BibitemShut
  {NoStop}%
\bibitem [{\citenamefont {Weber}(1961)}]{Weber}%
  \BibitemOpen
  \bibfield  {author} {\bibinfo {author} {\bibfnamefont {J.}~\bibnamefont
  {Weber}},\ }\href@noop {} {\emph {\bibinfo {title} {{General relativity and
  gravitational waves}}}}\ (\bibinfo  {publisher} {Interscience},\ \bibinfo
  {address} {New York},\ \bibinfo {year} {1961})\BibitemShut {NoStop}%
\bibitem [{\citenamefont {Maggiore}(2008)}]{Maggiore:2018sht}%
  \BibitemOpen
  \bibfield  {author} {\bibinfo {author} {\bibfnamefont {M.}~\bibnamefont
  {Maggiore}},\ }\href@noop {} {\emph {\bibinfo {title} {{Gravitational Waves:
  Volume 1: Theory and Experiments}}}}\ (\bibinfo  {publisher} {Oxford
  University Press},\ \bibinfo {address} {Oxford},\ \bibinfo {year}
  {2008})\BibitemShut {NoStop}%
\bibitem [{\citenamefont {Landau}\ and\ \citenamefont
  {Lifshitz}(1984)}]{LandauECM}%
  \BibitemOpen
  \bibfield  {author} {\bibinfo {author} {\bibfnamefont {L.~D.}\ \bibnamefont
  {Landau}}\ and\ \bibinfo {author} {\bibfnamefont {E.~M.}\ \bibnamefont
  {Lifshitz}},\ }\href
  {https://www.sciencedirect.com/book/9780080302751/electrodynamics-of-continuous-media}
  {\emph {\bibinfo {title} {{Electrodynamics of Continuous Media}}}}\ (\bibinfo
   {publisher} {Pergamon},\ \bibinfo {year} {1984})\BibitemShut {NoStop}%
\end{thebibliography}%

\end{document}